\documentclass[fleqn,10pt]{wlscirep}
\usepackage[utf8]{inputenc}
\usepackage[T1]{fontenc}
\usepackage{float}

\title{Eliminating nanometer-scale asperities on metallic thin films through plasma modification processes studied by molecular dynamics and AFM}%

\author[1,*]{Tomoyuki Tsuyama}%
\author[2]{Tatsuki Oyama}%
\author[2]{Yu Azuma}%
\author[2]{Haruhisa Ohashi}%
\author[2]{Masahiro Irie}%
\author[2]{Ayumi Yamakawa}%
\author[2]{Shoko Uetake}%
\author[2]{Takayuki Konno}%
\author[2]{Takahiro Ukai}%
\author[1]{Kohei Ochiai}%
\author[1]{Nobuyuki Iwaoka}%
\author[2,**]{Atsushi Hashimoto}%
\author[1]{Yoshishige Okuno}%

\affil[1]{Resonac Corporation, Research Center for Computational Science and Informatics \\ 8, Ebisu-cho, Kanagawa-ku, Yokohama, Kanagawa. 221-8517, Japan}%
\affil[2]{Resonac Hard Disk Corporation, Research \& Development Center \\ 5-1, Yawatakaigan-dori, Ichihara, Chiba, 290-0067, Japan}%

\affil[*]{tsuyama.tomoyuki.xixae@resonac.com}
\affil[**]{hashimoto.atsushi.xhtcs@resonac.com}

\begin{abstract}
  We report the effects of reducing surface asperity size at the nanometer scale on metallic surfaces by plasma-assisted surface modification processes using simulations and experiments. 
  
  Molecular dynamics (MD) simulations were conducted by irradiating various inert gas ions (Ne, Ar, Kr, and Xe) onto a cobalt slab with nanoscale asperities on the surface.
  The MD simulations showed that as the atomic number of the inert gas increased the surface asperity size was reduced more efficiently, while the etching rate decreased. 
  The dependencies of the scattering behaviors on the inert gas ions originated from the mass exchange between the working gas ions and the slab atoms.
  
  Atomic force microscopy and x-ray fluorescence measurements were performed on hard disk media subjected to the surface modification processes. 
  These measurements experimentally demonstrated that the density of nanoscale asperities was reduced with a lower etching rate as the atomic number of the inert gas increased, consistent with the simulation results.
  
  Through this study, we clarified that heavier working gases were more effective in reducing surface asperity size without significantly reducing the thickness of the material, which can contribute to better control of surface morphologies at the nanometer scale.

\end{abstract}

\begin{document}

\flushbottom
\maketitle

\thispagestyle{empty}

\section*{Introduction}
There has been a rapid increase in the demand for data storage over the past few decades, driven by digital transformation and critical technologies such as artificial intelligence and the Internet of Things \cite{Industry5, IDC_whitepaper}.
The amount of global data generated annually is expected to increase from  $\sim$ 16.1 zettabytes to $\sim$ 163 zettabytes per year through 2016 to 2025. 
Among various types of storage devices, hard disk drives (HDDs) serve as the primary storage devices in data centers, because of their low cost and huge capacities \cite{IDC_whitepaper}. 
The continuous improvements in the capacity and reliability of HDDs have supported innovations in digital technologies in global society \cite{hamr_technology_roadmap, technology_roadmap, hamr_fept_overview}.

The reliability of HDDs is of paramount importance to ensure the secure long-term storage of data \cite{hdi_roadmap}.
Numerous studies have explored the correlation between the reliability of HDDs and their surface morphologies. 
These studies have claimed that nanoscale asperities on the surface of hard disk media can cause critical damage to the read/write heads \cite{hdi_roadmap, reliability_modelling, roughness_effect, fem_sperity}.
Therefore, reducing the size and density of these asperities is a crucial process for improving the reliability of HDDs.

However, heat-assisted magnetic recording (HAMR) media, which have been extensively developed to achieve breakthroughs in the capacity as the next generation of HDDs, have a much higher density of nano-asperities compared to conventional magnetic recording (CMR) systems \cite{roughness_HAMR_MAMR_comparisons}. 
This is because of the difficulties in controlling morphologies, such as rough surfaces in the underlayers of MgO \cite{rough_MgO}, secondary particles during FePt thin-film growth \cite{secondary_particle}, and the inhomogeneous heights of FePt nano-grains \cite{BN}.

Therefore, reducing the size and density of asperities on hard disk media is critical for the reliability not only for current-generation HDDs (CMR) but also for next-generation HDDs (HAMR). 
To reduce the size and density of asperities, surface modification processes using Ar gas are known to be effective \cite{etching_remove_asperity1, etching_remove_asperity2}. 
However, the atomic-scale mechanisms in the smoothing of surface asperities and strategies for further improvements are yet to be fully understood.

In this paper, we report the effect of ion bombardments on nano-asperity through molecular dynamics (MD) simulations and surface modification processes applied to HDD media experimentally. 
The simulations and experiments consistently showed that the heavier the atomic mass of the working gas is, the more effective it is at reducing asperity size, despite the lower etching rate. 
The dependence of gas species to reduce asperity sizes originates from the atomic mass affecting the exchanges of momenta between the inert gas and the surface atoms.

\section*{Methods}
We employed MD simulations to understand the processes involved in eliminating asperities from metallic surfaces through ion impacts at the atomic scale. 
To verify the results of simulations, hard disk media subjected to surface modification processes with various inert gas species were measured by atomic force microscopy (AFM) and x-ray fluorescence (XRF) measurements.
This section provides detailed information about the conditions of simulations and experiments.

\subsection*{MD simulations}
We used the Atomic Simulation Environment (ASE)\cite{ASE} and the Large-scale Atomic/Molecular Massively Parallel Simulator (LAMMPS)\cite{LAMMPS} for pre- and post-processing, as well as for the MD simulations respectively. OVITO and VESTA were employed for analysis through visualization\cite{OVITO, VESTA}.

\begin{figure}[H]
\begin{center}
  \includegraphics[height=0.2\textheight]{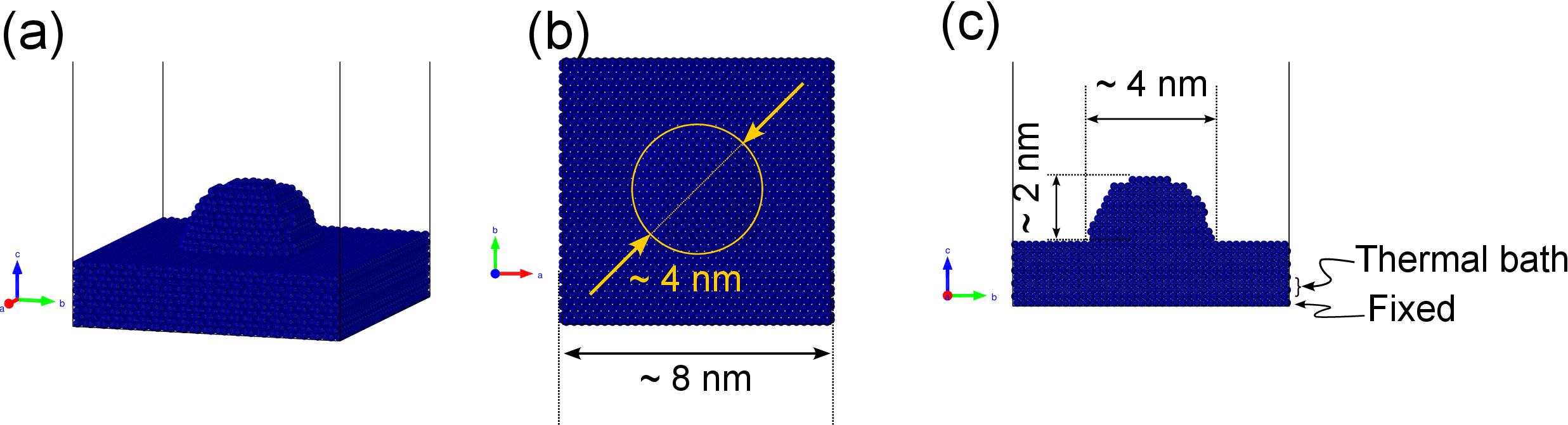}
  \caption{The initial structures of slabs simulated in this paper are illustrated, from the standard orientation of crystal shape (a), and top (b), and side (c) view. A semi-spherical asperity is artificially set on a slab. They are composed of 16000 of Cobalt (Co) atoms in total.}
  \label{fig:slab}
\end{center}
\end{figure}

To study the surface modification process of nanoscale asperities on metallic surfaces, we constructed a Co slab with an artificial asperity as shown in Figure \ref{fig:slab}. 
Co alloy is commonly used as the top layer in CMR and HAMR\cite{PMR_cap, HAMR_cap}. 
Figure \ref{fig:slab} (a) - (c) illustrates the initial slab used in this simulation. A flat Co slab with a thickness of 2 nm and a periodic structure in a square shape with sides of $\sim$ 8 nm was prepared as the base substrate. At the center of the thin-film slab, we introduced a Co protrusion with a semi-spherical shape and a radius of $\sim$ 2 nm\cite{fem_sperity}. The slab with the protrusion consists of about 16,000 Co atoms in total.

During the simulation, the regions are divided into three, as explained in Figure \ref{fig:slab} (c). 
The bottom layer is frozen to maintain the total momentum of the entire cell at zero during collisions of ions and to prevent it from drifting. 
The $X$ and $Y$ directions are set as periodic dimensions, while the $Z$ direction is set as non-periodic. 
The temperature of the middle layer of the thin film is maintained at 700 K using a Berendsen thermostat\cite{berendsen}, which operates every 100 femtoseconds\cite{FePt_etching, FePt_etching_reactive, etching_Tinacba_2021, etching_Joseph_2005, etching_Satish_1995}.

Two types of interatomic potentials are employed. 
Modified embedded-atom method (MEAM) potentials are applied among Co atoms, which describe the various physical properties of bulk and surfaces accurately \cite{meam, meam_Co}.
The accurate descriptions of the surface binding energy of the slab by MEAM potentials are the key advantages of MD to simulate the atomic scattering at surfaces, especially when the energy of incident ions is less than 1 keV \cite{MD_vs_BCA}.
The interactions between Co and inert ions were computed using Ziegler-Biersack-Littmark (ZBL) potentials to account for strong repulsions during high-energy collisions of atoms \cite{zbl, FePt_etching}. 

Ar$^+$ ions are treated in the simulation as neutral Ar atoms. 
This is because the charge of the injected ions is immediately compensated by Auger emission through interactions with the surface \cite{neutralization_by_auger, neutralization_by_auger2}. 
After ion injection,the atomic coordinates of the system (i.e., surface material and injected ions) evolve under fixed total energy (i.e., microcanonical). 
Energy is gradually transferred to the heat bath beneath the surface layer, utilizing the Berendsen thermostat.

Ion injection occurs every 6 picoseconds, corresponding to the ion fluxes of the order $10^{23} \text{cm}^{-2}\text{s}^{-1}$, which is significantly higher than that of our experiment ($10^{15} \text{cm}^{-2}\text{s}^{-1}$).
This high ion flux is necessary because MD simulations require reasonable computational speed, and therefore, the irradiation rate in the simulations is typically set much higher than that in experiments \cite{FePt_etching, rate_example, rate_example2}.
The interval time for the ion insertion of 6 picoseconds is determined by the relaxation time for target thermalization to 600 K (see Figure S1, S2, and Table S1 in Supplementary Materials)\cite{Dominguez-Gutierrez_2017, Bedoya2019}.

The time step size in this simulation is set to 1 femtosecond. 
Ion injections at normal incidence ($-Z$ direction) are initiated from a 3D box source placed above the surface. 
The initial velocity of an ion is given by $v = \sqrt{\frac{2qV}{m}}$
where $q$ denotes elementary charge, $m$ is the atomic mass of injected ions, and $V$, applied bias, is varied through 50, 100, 150, 200 and 250 V. 
Ejected particles are allowed to exit the simulation space so that the number of atoms remained in the simulation cell and the number of atoms sputtered off can both be counted. 

\subsection*{Description of experimental}

A $\sim$7 nm thick layer of Fe$_{50}$Pt$_{50}$ $L1_0$ grains (FePt) is grown on MgO seed layers above adhesion and heat sink layers using Anelva sputtering equipment at a temperature of $\sim$700 K. 
On top of the FePt grains, Co is deposited following a 6-second surface modification process, which is applied using an RF bias (13.56 MHz) while maintaining a working pressure of 2.0 Pa. We compare the morphologies using Ar, Kr, and Xe as the working gases to verify their dependencies on the gas species \cite{Dwivedi2015}.

Finally, 1.5 nm carbon overcoats (COC) on media are deposited for tribological, corrosion-resistant and oxidation-resistant characteristics.

To characterize the surface morphologies we paid attention to kurtosis ($R_{\textrm{Ku}}$) obtained by AFM measurements (Park NX-HDM) with tapping mode.
$R_{\textrm{Ku}}$, a statistical measure of the "tailedness" of the probability distribution of a real-valued random variable, is particularly well-suited for characterizing the density of nanoscale protrusions. 
In AFM analysis, $R_{\textrm{Ku}}$ is derived from the fourth central moment of the surface height distribution, mathematically represented as:
$$
R_{\textrm{Ku}} = \frac{\frac{1}{n}\Sigma_{i=1}^{n}(x_i - \bar{x})^{4}}{\left[\frac{1}{n}\Sigma_{i=1}^{n}(x_i - \bar{x})^{2}\right]^2}
$$
where $x_{i}$, $\bar{x}$, and n denote $\textit{i}^{th}$ height data, the mean height of $x_{i}$, and the number of the data over the measurement area, respectively.
A higher $R_{\textrm{Ku}}$ value ($R_{\textrm{Ku}}$ > 3) indicates that the surface exhibits more frequent and severe deviations from the mean height. Therefore, $R_{\textrm{Ku}}$ serves as an effective index for quantifying the density of protrusions.

XRF measurements (Rigaku AZX 400) were performed to evaluate the thickness of the Co layer by sweeping the Co $K_{\alpha}$ intensities, which were calibrated using transmission electron microscopy.

\section*{Simulation results}

In this section, we investigate the dependence of surface modifications on various gas species in reducing surface protrusion size and etching rates through MD simulations.
\subsection*{Effect of continuous ion bombardments on a metallic slab}

\begin{figure}[H]
\begin{center}
\includegraphics[height=0.2\textheight]{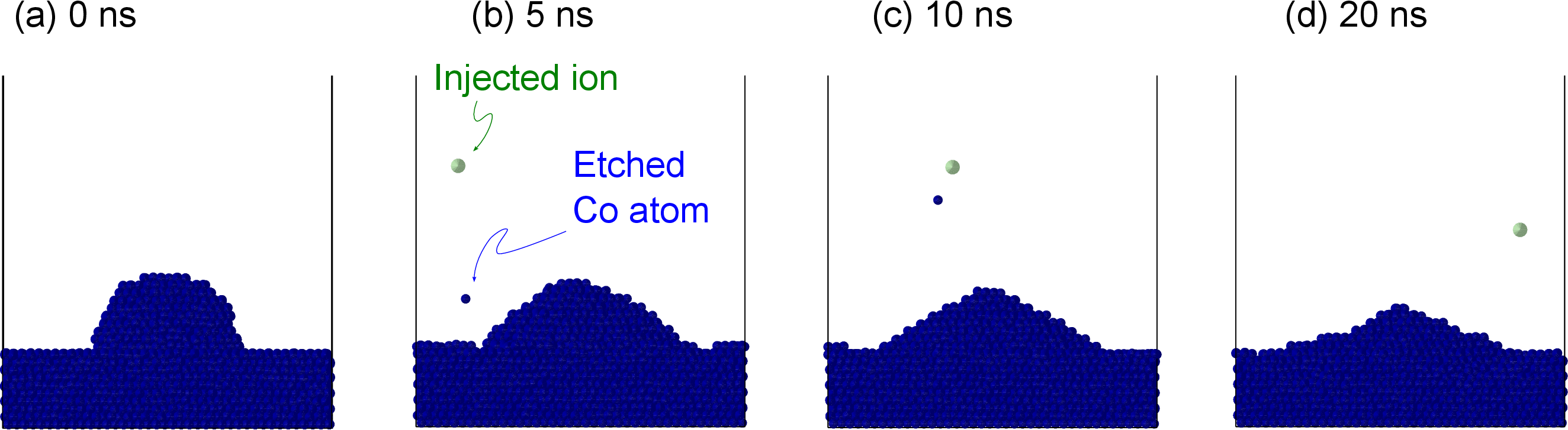}
\caption{
Snapshots of the etched slab taken at 0 (a), 5 (b), 10 (c), and 20 (d) nanoseconds from the MD simulation, with a 100 V acceleration applied to Ar ions.
The size of the protrusion is gradually decreased, and the Co atoms in the slab etched away by the bombardments of inert-gas ions.
}
\label{fig:snapshots}
\end{center}
\end{figure}

Figure \ref{fig:snapshots} illustrates snapshots from the MD simulations, using Ar with 100 V of bias as an example.
The simulation movies were prepared in supplementary materials for reference.
The snapshots in Figure \ref{fig:snapshots}(a) through (d) were captured at 0, 5, 10, and 20 nanoseconds of elapsed time. 
These snapshots qualitatively demonstrate that the size of the asperity decreases as more inert gas ions are introduced, resulting in the etching and removal of a part of Co atoms from the simulation cells.
Movie files of the MD simulations are available as Supplementary Materials.

\begin{figure}[H]
\begin{center}
\includegraphics[width=0.8\linewidth]{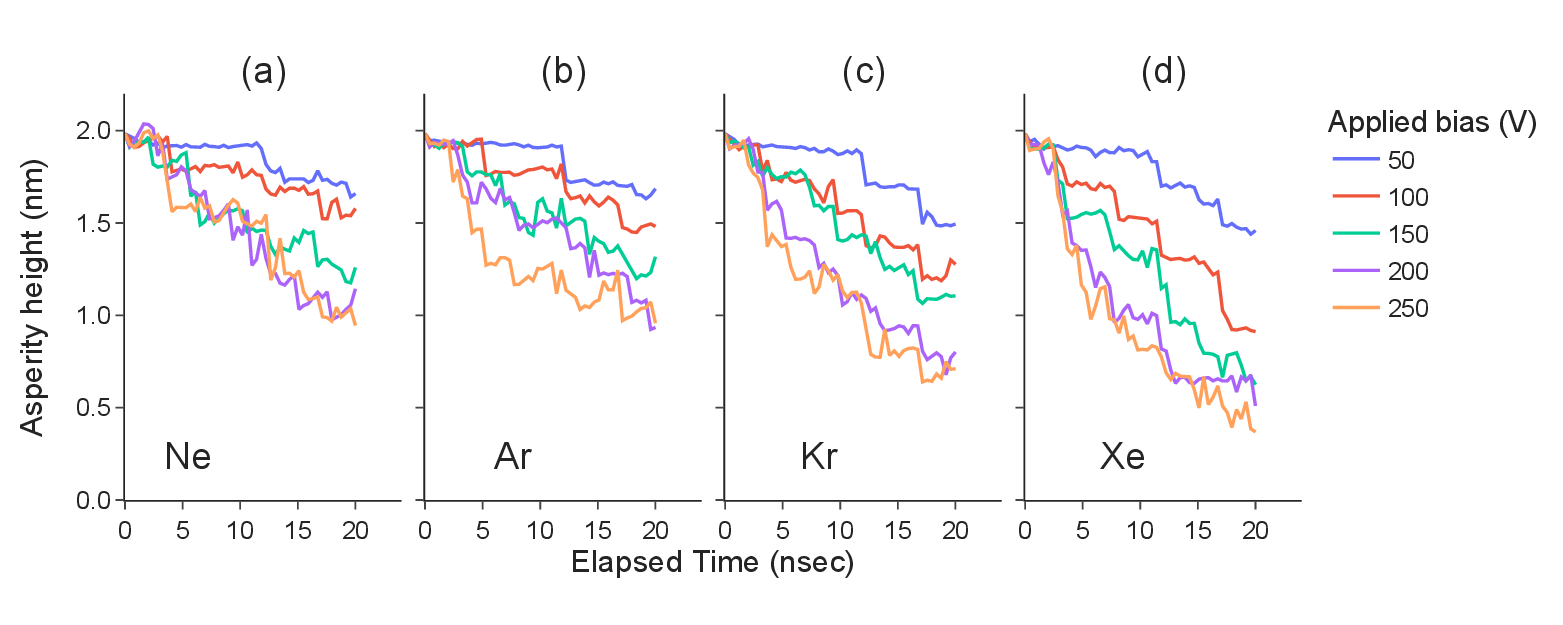}
\caption{The residual height of the asperity on the slab during surface modification process with various inert gases: (a) Ne, (b) Ar, (c) Kr, and (d) Xe. To illustrate the dependence on the applied bias, the MD simulation results under different bias conditions are represented in distinct colors. The definition of the height of the asperity is provided in the main text.}
\label{fig:asperity_vs_gas}
\end{center}
\end{figure}

Figure \ref{fig:asperity_vs_gas} presents the height of the asperity obtained at each time step for various ingredient gas species: (a) Ne, (b) Ar, (c) Kr, and (d) Xe.
The height of the asperity is defined as the distance between the peak of the asperity and the mean displacement at the top of the slab along the $Z$-axis.
The $Z$-value of the apex of the asperity is determined by taking the average of the $Z$-values of the highest 5\% of the atoms constituting the asperity.
As qualitatively illustrated in Figure \ref{fig:snapshots} (a) through (d), the height of the asperity decreases gradually with surface modification.
A stronger applied bias reduces the size of the asperity more effectively because the ions are more strongly accelerated and possess greater kinetic energy.
We also found that the asperity can be reduced more significantly with higher atomic number ions.

\begin{figure}[H]
  \begin{center}
  \includegraphics[width=0.8\linewidth]{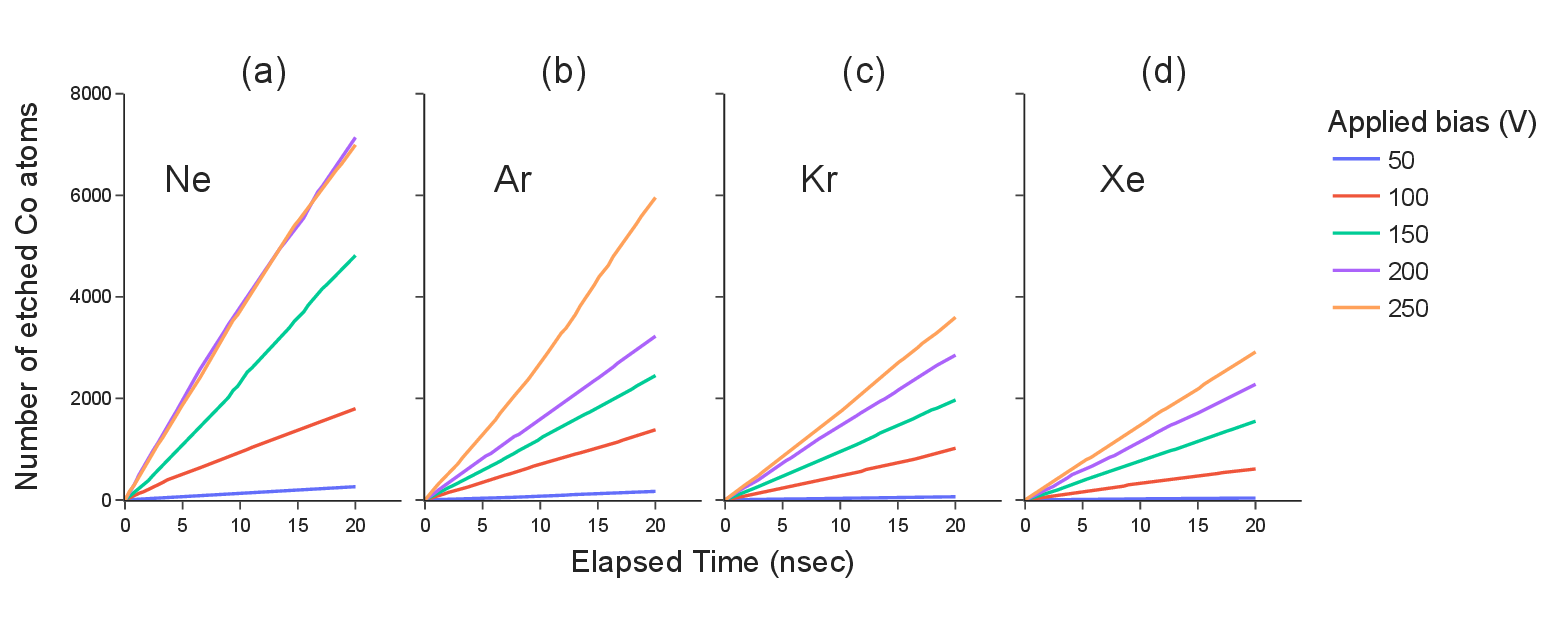}
  \caption{The number of etched Co atoms as a function of elapsed time for various inert gas species: (a) Ne, (b) Ar, (c) Kr, and (d) Xe.}
  \label{fig:etched_atoms_vs_gas}
  \end{center}
\end{figure}

Figure \ref{fig:etched_atoms_vs_gas} depicts the number of Co atoms etched away from the simulation cell. 
The number of etched Co atoms increases monotonically with elapsed time and applied bias, owing to the higher energy transferred from the kinetic energy of the injected ions. Among the tested working gases, we observed that ions with higher atomic numbers exhibited lower etching rates.
This tendency is consistent with previous studies investigating the etching rates of metallic and semiconductor surfaces \cite{etch_rate_segregants, etch_rate_semiconductor}.

To analyze this data numerically, we introduce the expressions of sputter yield advocated by Sigmund \cite{Sigmund1969}.
When an ion with the kinetic energy of $V$ and with the mass of $M_i$ is injected into the slab with the atomic mass of $M_s$, the sputtering yield $Y$ is expressed by:
$$
Y \approx \dfrac{3}{4\pi^2}\dfrac{\alpha \Lambda V}{E_{sb}} .
$$
Where $\Lambda = \dfrac{4M_sM_i}{(M_s + M_i)^2}$ is the energy transfer factor, $\alpha$ is the dimensionless factor, and $E_{sb}$ is the surface binding energy of slabs.
This is valid only when the energy of the incident ions $V$ is significantly greater than the surface binding energy $E_{sb}$, due to the characteristics of the linear cascade regime, and is derived from several simplifications applicable to ion energies below 1 keV.
Mahne $et al$. have clarified that $\alpha$ can be approximated by $0.15(1 + M_s/M_i)$ \cite{Mahne2022}.
As a result, the expression for the sputter yield can be simplified to:
$$
Y \approx \dfrac{0.45}{\pi^2}\dfrac{M_s}{M_s + M_i}\dfrac{V}{E_{sb}} .
$$
Therefore, the sputter yield is proportional to $\dfrac{M_s}{(M_s + M_i)}$, and is decreased with respect to $M_i$, which explains the gas ion dependencies of the etching rate shown in Figure \ref{fig:etched_atoms_vs_gas} (see Figure S5 in Supplementary Materials for the comparisons with the actual sputter yield).

\begin{figure}[H]
\begin{center}
\includegraphics[width=0.7\linewidth]{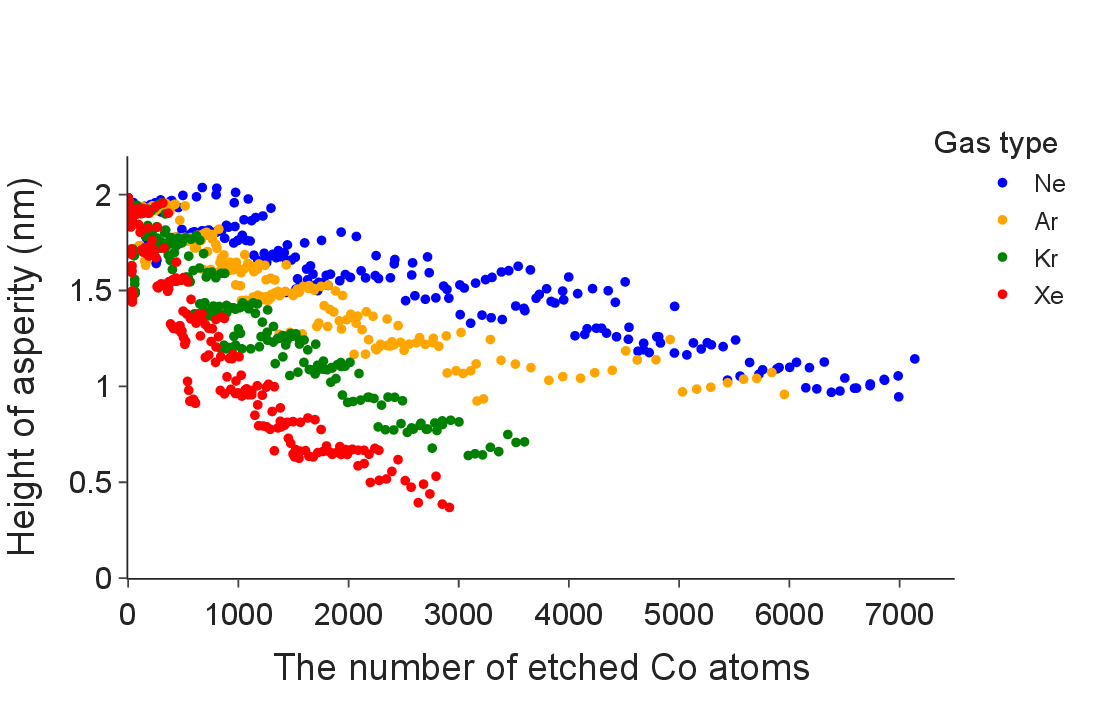}
\caption{Plots for the residual height of the asperity and the number of etched atoms during the ion bombardments with various types of inert gases.}
\label{fig:etched_atoms_vs_asperity}
\end{center}
\end{figure}

Figure \ref{fig:etched_atoms_vs_asperity} depicts the number of etched Co atoms and the height of the asperity for each time step, combining data from Figure \ref{fig:asperity_vs_gas} and Figure \ref{fig:etched_atoms_vs_gas}. 
Each gas is plotted in distinct zones, making the type of gas a unique parameter for controlling both the final surface morphology and the residual film thickness of the thin film; these factors cannot be controlled independently by substrate bias alone. 
From this plot, we conclude that surface modification with higher atomic number gas can reduce the height of asperities more efficiently while suppressing the reduction of film thickness.

\subsection*{Behavior of single ion bombardment}
To understand the mechanism of surface redistribution of metal atoms through ion bombardments depending on inert gas species, we conducted a more detailed analysis. We injected a single ion onto the Co slab and observed the scattering using various types of gases.

\begin{figure}[H]
\begin{center}
\includegraphics[height=0.3\textheight]{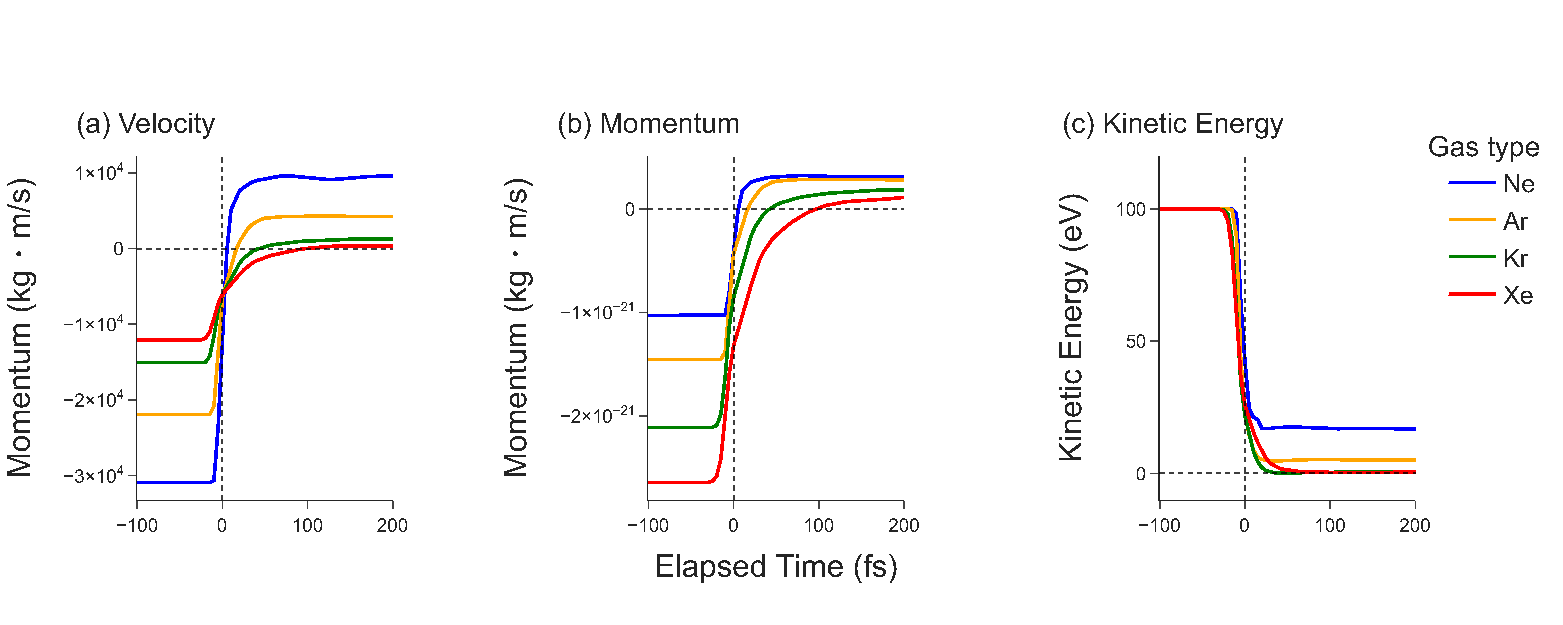}
\caption{The $Z$-components of velocities (a), momenta (b), and kinetic energies (c) of a single inert gas ion with applied biases of 100 V are plotted before and after collision with the slab surface. 
}
\label{fig:velocity_z}
\end{center}
\end{figure}

Figure \ref{fig:velocity_z} compares the time evolution of the $Z$-component in (a) velocities, (b) in momenta, and (c) kinetic energies for various tested gases. 
Time zero is set to the moment of collision between the inert gas ion and the Co slab, determined by the nearest distances between the ion gas and slab atoms. We repeated this simulation 10 times with different seeds and averaged the extracted values.
We varied the bias to 50, 100, 150, 200, and 250 V, but only 100 V is displayed in Figure \ref{fig:velocity_z} because there were no significant differences in conclusions among these changes.
For $\textit{t}$ < 0, both the velocities and momenta are negative due to the initial velocities.
After the collision at $\textit{t}$ = 0, the velocities, momenta, and kinetic energies suddenly change toward zero or even reverse sign, implying that the ion is reflected at the surface of the slab. 

The sign of velocity and momenta changes more slowly for the higher atomic number ions, indicating that these ions are scattered forward after their first collision. In contrast, the smaller atomic number gases change their velocities more rapidly after the collision, implying back-scattering. 
As a result, the heavier ions interact with more atoms by passing the momenta of the gas ion to slab atoms.

The kinetic energies of inert gas ions after the collisions shown in Figure \ref{fig:velocity_z} (c) decreases monotonically as the masses of the inserted ion increases. 
However, the transferrable energies in a single collision, $\Lambda$, do not increase monotonically with respect to $M_i$. 
For example, the $\Lambda$ value of Xe is lower than that of Ar as shown in Figure S6 in Supplementary Materials. 
Therefore, the heavier ions should collide with a greater number of atoms to explain the systematically lower kinetic energy after collision with $M_i$.

The differences in scattering by the gas species can be qualitatively explained by the classical mechanics of collisions between two rigid bodies. 
Because of the forward scattering in higher atomic number ions, they interact with more atoms by digging into the slab, which promotes the redistributions of atomic positions. 
On the other hand, the lighter ions are back scattered at the surface, by giving a strong impulse to fewer surface atoms, close to a single knock-on regime \cite{Eckstein2007}. Because lighter ions pass their moment locally within a shorter time, it can induce a higher sputtering yield.


\section*{Experiment results}
This section experimentally examines how the ingredient gas affects the removal of surface protrusions and the etching rate.
\subsection*{AFM measurements}

\begin{figure}[H]
\begin{center}
\includegraphics[width=0.8\linewidth]{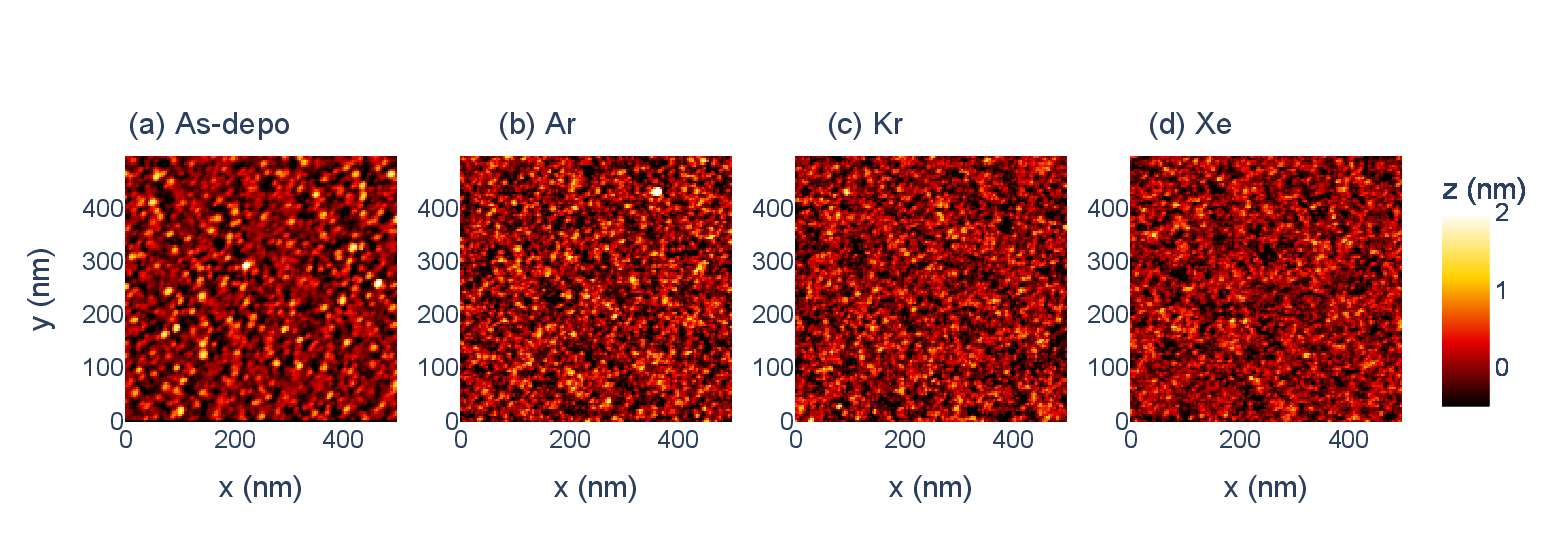}
\caption{\label{fig:afm} AFM images of hard disk media measured on COC: (a) as-deposited, (b) etched by Ar, (c) etched by Kr, and (d) etched by Xe, with a substrate RF bias power of 200 W applied to the specimens.}
\end{center}
\end{figure}

Figure \ref{fig:afm} displays the AFM images measured on the carbon protection layer used to prevent oxidation, for the samples without surface modification process (a), and with the process using Ar (b), Kr (c), and Xe (d), applying a bias power of 200 W.
The density of protrusions at the nanoscale (white spots in the AFM images) is high without the surface modification process, and it gradually decreases when applying the process with Ar, Kr, and Xe.

To quantitatively evaluate the effect of the surface modification process on the capability of removing surface asperities, we extracted the statistical values from AFM images (Figure \ref{fig:afm}) in Table \ref{tab:afm_parameters}.
We compare the density of surface protrusions using the parameter $R_{\textrm{Ku}}$ as defined in the previous section, obtained from AFM images ($N = 3$), which are treated by the surface modification process with a bias power of 200 W.
Although we were unable to observe a clear trend in $RMS$, $R_{\textrm{Ku}}$ has a systematic trend: $R_{\textrm{Ku}}$ can be reduced through the surface modification process, and this reduction is achieved more efficiently when using gas species with higher atomic numbers.
This trend is consistent with the MD simulations demonstrated in the previous section.

\subsection*{Etching rate}
\begin{figure}[H]
  \begin{center}
  \includegraphics[height=0.3\textheight]{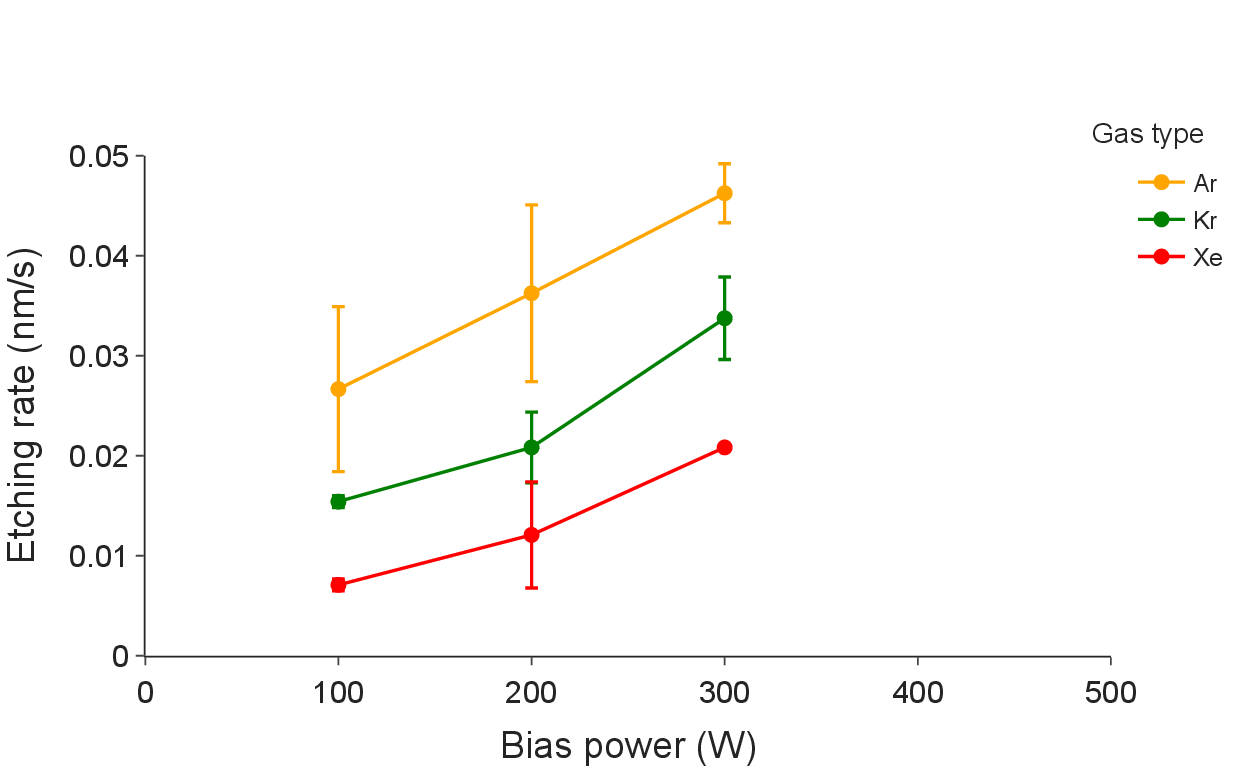}
  \caption{\label{fig:etching_rate} The substrate bias-power dependence of etching rate for various ingredient gases.}
  \end{center}
\end{figure}

To determine the etching rate, Co alloy is deposited onto an aluminum substrate with a thickness of $\sim$ 2 nm. Subsequently, 6 seconds of surface modification process is executed with various bias powers to be etched.
We acquired the etching rate $(T_{\textrm{etch}} - T_{\textrm{as-depo}}) / t $.
Where, $T_{\textrm{etch}}$, $T_{\textrm{as-depo}}$, and $t$ denote the thickness of Co after, before etching, and the process duration time.
The etching rates with various inert gases are displayed in Figure \ref{fig:etching_rate}.
The etching rate increases with higher substrate biases power and with gases of smaller atomic numbers, consistent with MD simulations.


\section*{Conclusion}
In conclusion, we investigated the mechanisms of plasma modification on metallic surfaces using inert gases, focusing on the capability to remove surface asperities through classical MD simulations and experiments (AFM and XRF).

\begin{itemize}
\item MD simulations demonstrated higher efficiency in removing Co asperities from surfaces using higher atomic number etching gases, such as Xe, although the etching rate is lower.
\item The differences in surface modifications driven by gas species are attributed to the mass of the gas ions, which influences the exchange of momentum at the moment of collisions.
\item AFM measurements on hard disk media, after applying the surface modification process, suggested that it is a powerful method to reduce surface protrusion size, especially when high atomic number working gases are used. XRF measurements also revealed that the etching rate of higher atomic number inert gases is lower. These experimental observations are consistent with the simulations.
\end{itemize}

To decrease asperity size on the surface, surface modification processes using high atomic number gases, such as Xe are a better choice for efficiency. 
The surface modification process using such heavier gases is likely to be helpful for HDD media to reduce surface asperity without decreasing the thickness of the Co layer deposited on the top of magnetic layers.
On the other hand, for cleaning or removing segregants from metallic surfaces, working gases with smaller atomic numbers can achieve this more efficiently, based on our study. 
Our research could contribute to better control of surface morphology on metals, not only for hard disk media but also for general applications at the nanometer scale.

\section*{Data Availability}
All data generated or analyzed during this study are included in this published article.
If required, any data are available from the corresponding author on reasonable request.

\nocite{*}
\bibliography{reference}

\section*{Author contributions statement}
T.T and A.H supervised this work.
T.T, K.O, and N.I performed the simulations.
T.O, Y.A, H.O, H.I, T.U, and A.H designed the experiments.
T.T, T.O, H.O, A.Y, S.U, and T.K analyzed the data.
T.T wrote the main manuscript text and prepared the figures.
All authors reviewed the manuscript.

\section*{Additional Information}
The authors declare no competing interests.

\begin{table}[H]
  \centering
  \caption{\label{tab:afm_parameters} Table to summarize the parameters obtained by AFM measurements on hard disk surfaces without the surface modification process (as-depo), with the process by Ar, Kr and Xe.}
  \resizebox{0.4\textwidth}{!}{
  \begin{tabular}{lccc}
    & $RMS$ (nm) & $R_{\textrm{Ku}}$\\ \hline
  As-depo & 0.42 $\pm$ 0.04 & 5.1 $\pm$ 0.3\\
  Ar & 0.49 $\pm$ 0.03 & 4.0 $\pm$ 0.2\\
  Kr & 0.46 $\pm$ 0.03 & 3.3 $\pm$ 0.1\\
  Xe & 0.42 $\pm$ 0.03 & 3.1 $\pm$ 0.1\\
  \end{tabular}
  }
\end{table}

\end{document}